\documentclass[preprint,,showpacs,preprintnumbers,amsmath,amssymb]{revtex4-1}

\usepackage{graphicx}
\usepackage{epstopdf}
\usepackage{dcolumn}
\usepackage{bm}
\usepackage{amsmath}
 
\begin{document}

\title{Coherent Band-Edge Oscillations and Dynamic LO Phonon Mode Splitting as Evidence for Polaronic Coupling in Perovskites 
} 

\author
{Z. Liu$^{1}$, C. Vaswani$^{1}$, L. Luo$^{1}$, D. Cheng$^{1}$, X. Yang$^{1}$, X. Zhao$^{1}$, Y. Yao$^{1}$, 
	Z. Song$^{2}$, R. Brenes$^{3}$, R. Kim$^{1}$, J. Jean$^{3}$, V. Bulovi$\acute{c}^{3}$, Y. Yan$^{2}$, K.-M. Ho$^{1}$,
	J. Wang$^{1\ast}$
	\\
	\normalsize{$^{1}$Department of Physics and Astronomy, Ames Laboratory, Iowa State University, Ames, IA 50011 USA}\\
	\normalsize{$^{2}$Department of Physics and Astronomy and Wright Center for Photovoltaics Innovation and Commercialization, The University of Toledo, Toledo, OH 43606, USA}\\
	\normalsize{$^{3}$Department of Electrical Engineering and Computer Science,}\\
	\normalsize{ Massachusetts Institute of Technology, Cambridge, MA 02139, USA}\\
}

\date{\today}

\begin{abstract}
The coherence of collective modes, such as phonons, and their modulation of the electronic states are long sought in complex systems, which is a cross-cutting issue in photovoltaics and quantum electronics. In photovoltaic cells and lasers based on metal halide perovskites, the presence of polaronic coupling, i.e., photocarriers dressed by the macroscopic motion of charged lattice, assisted by terahertz (THz) longitudinal optical (LO) phonons, has been intensely studied yet still debated. 
This may be key for explaining the remarkable properties of the perovskite materials, e.g., defect tolerance, long charge lifetimes and diffusion length. 
Here we use the intense single-cycle THz pulse with the peak electric field up to $E_{THz}=$1000\,kV/cm to drive coherent band-edge oscillations at room temperature in CH$_3$NH$_3$PbI$_3$.  
We reveal the oscillatory behavior dominantly to a specific quantized lattice vibration mode at $\omega_{\mathrm{LO}}\sim$4 THz, being both dipole and momentum forbidden. 
THz-driven coherent dynamics exhibits distinguishing features: the room temperature coherent oscillations at $\omega_{\mathrm{LO}}$ longer than 1 ps in both single crystals and thin films; the {\em mode-selective} modulation of different band edge states assisted by electron-phonon ($e$-$ph$) interaction; {\em dynamic mode splitting} controlled by temperature due to entropy and anharmonicity of organic cations. 
Our results demonstrate intense THz-driven coherent band-edge modulation as a powerful probe of electron-lattice coupling phenomena and provide compelling implications for polaron correlations in perovskites.     
\end{abstract}

\maketitle
\preprint{Manuscript to PRB}

A challenge underlying efficient photoenergy conversion is to protect photocarrier transport from scattering with disorders. These include static disorders, such as defects and grain boundaries, or dynamic ones \cite{Miyata,Fu,Prakriti,Kiyoshi,Micheal,zhu,Chen,Fan}, such as phonons, rapid organic cation fluctuations and multicarrier Auger excitations\cite{Liu,Luo,Gary}. 
Perovskites has been identified as a model system for understanding and implementing the large {\em ferroelectric polaron} formation mechanism\cite{Miyata,Fu,Kiyoshi} that screens Coulomb potential of charge carriers 
for disorder-immune photoconversion and slow electron-hole ($e$-$h$) recombination, among other attractive properties \cite{zhu,Chen,Fan}.
In additional to the generic polaronic coupling, or 
charge coupling to phonons, ubiquitous in polar lattice\cite{book1}, the large ferroelectric polarons \cite{Miyata} in perovskites are very relevant for shielding photocarrier transport. The mechanism has been discussed \cite{Kiyoshi,Miyata} in the context of the polarization cloud assisted by LO phonon-electron interaction, as illustrated in Fig. 1a. For example, in analogy to ferroelectric materials, the high frequency LO mode at $\omega_{\mathrm{LO}}\sim$4 THz \cite{Nagai}, both optically dark and momentum forbidden, has been discussed to provide the macroscopic motion of charged lattice and internal electric field to mediate the large polaron formation. 
Previous studies have gained important insights into $e$-phonon interaction and polaronic couping in perovskites using Raman scattering\cite{Omer} and coherent phonon pumping\cite{Gio,Tufan,Chengbin,Thouin}. However, the high frequency LO modes $\sim$4 THz thus far is much less observed by ultrafast optical studies, which reveals mostly lower frequency modes \cite{Chengbin,Thouin}.   
Recently, the high-frequency LO phonon dynamics $\sim$3.7 THz is observed via optical pump and weak THz probe at room temperature \cite{McGill}. 
The next major challenge is to explore the LO mode-splitting, controlled by temperature, especially for the 4 THz LO photon mode. 
In addition, it has not been revealed {\em$\omega_{\mathrm{LO}}$ mode-specific} coherent coupling to the band-edge states relevant for optoelectronic modulator and device applications. 

\begin{figure*}[!tbp]
	\includegraphics[scale=0.6]{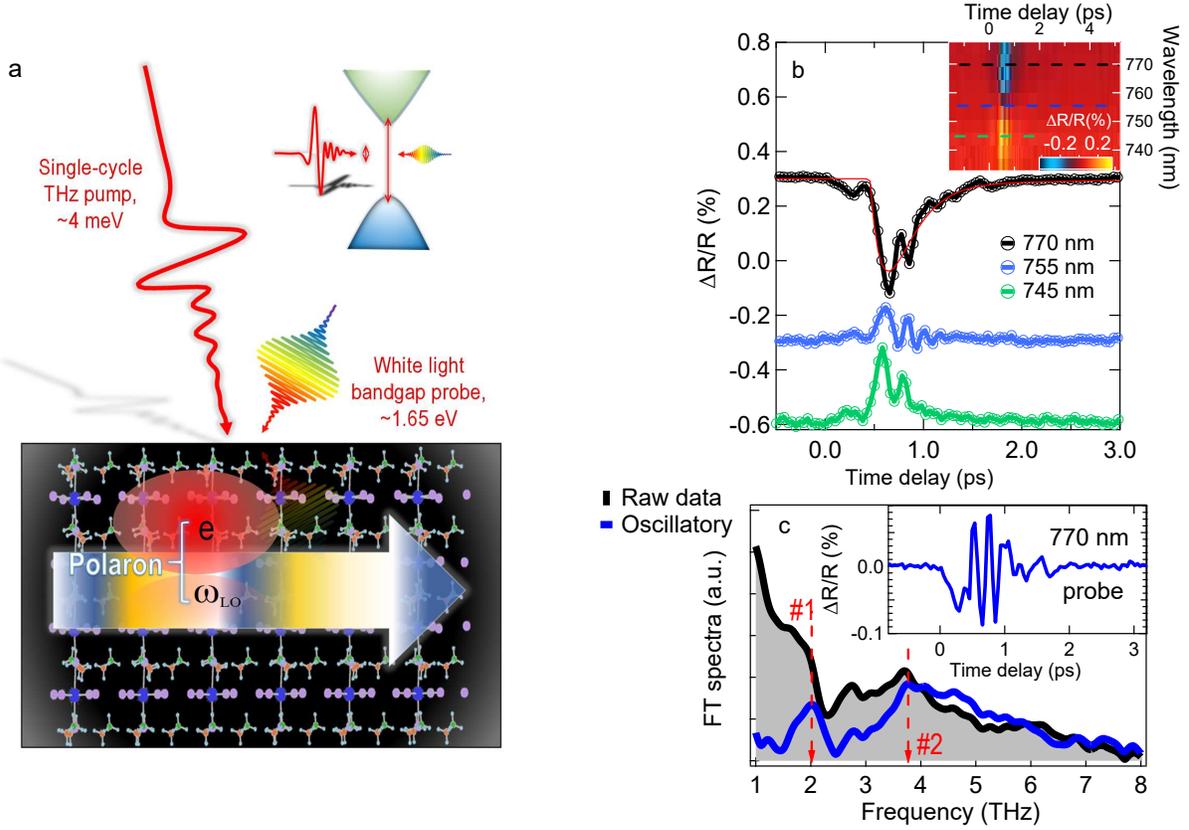}
	\caption{\textbf{Room temperature coherent band-edge dynamics in MAPbI$_3$ driven by the intense single-cycle THz electric field pulse.} (a) An illustration of polaron formation in MAPbI$_3$ after THz pumping. The $\omega_{LO}$-model specific coupling to band edge states can be probed by a broadband white light continuum (inset, top left corner). (b) Three selective $\Delta$R/R time traces at probe wavelength of 770 nm (black), 755 nm (blue), and 745 nm (green) induced by THz pump electric field E$_{THz}$ = 938 kV/cm at room temperature. The red trace shows the exponential decay components at 770 nm obtained from fitting (methods). Inset: 
		A two-dimensional false-colour plot of the full transient spectra as a function of time delay $\Delta$t$_{pp}$. The dashed lines marked the wavelength cut of three $\Delta$R/R dynamics selected. (c) The Fourier transform (FT) spectra of the $\Delta$R/R (black) and oscillatory component $\Delta$R$_{osc}$/R (blue) at 770 nm.  Two pronounced peaks centered at $\sim$4 THz and $\sim$2 THz are marked as red dashed lines. Inset: raw $\Delta$R$_{osc}$/R data in time domain.}
	\label{Fig1}
\end{figure*}

Probing coherent band-edge modulation, assisted by the LO phonon $\omega_{LO}$, and its selective electronic coupling to different bands requires overcoming the challenges of detecting dynamics of the $\omega_{\mathrm{LO}}$ mode, an optically-dark one at higher center-of-mass momentum $K$-space. They are restricted from commonly-used interband optical transitions. 
Yet, the challenge can be met by coherent excitation and detection methods driven by an intense single-cycle THz pulse at fs time and THz frequency scales. 
As illustrated in Fig. 1a, subcycle acceleration driven by intense THz electric field expects to 
impulsively distort high-frequency vibrational potential of LO phonons by dynamically generating internal electric field in MAPbI$_3$. 
For example, THz-induced interband coherence has been identified to assist tunneling of electrons in GaAs \cite{GaAs}. 
The soft nature of organic-inorganic hybrid perovskite materials make the ions reside in very shallow minima of the potential energy landscape and can strongly couple to band edge electronic states unlike the GaAs.  
In addition, the coherent LO phonons can have influence on electronic bands depending on their symmetry-dependent, selective-coupling to the $\omega_{\mathrm{LO}}$ mode assisted by the polaronic coupling.  
However, the LO-phonon-induced band-edge modulation is not observed in prior THz pumping measurements\cite{Kim} using relatively low electric field ($\sim$100 kV/cm) 
that focus on an optical-active, transverse optical phonon, i.e., TO mode, at low frequency $\sim$1 THz 
A megavolt/cm THz pumping electric field use here, one order of magnitude higher than prior studies, is required for exciting $\omega_{\mathrm{LO}}$ coherence and band-edge modulation. 

Here we reveal the robust $\omega_{\mathrm{LO}}$ quantum oscillation driven by THz pumping at room temperature in both single crystal and thin film MAPbI$_3$ samples. The intense, sub-picosecond single-cycle THz pulse\cite{Xu,Xu1, Liang1} has the peak electric field up to $E_{THz}=$1000\,kV/cm and deep-subgap photon energy centered at $\omega_{\mathrm{pump}}\sim$1.5 THz or 6.2 meV. This approach allows the observation of a LO mode-selective coupling to different band edge states.  
We also reveal LO phonon coherent dynamics, from mode splitting to beating, controlled by temperature due entropy and anharmonicity of the organic cations.   
Intense THz-driven quantum beats demonstrated not only represent a powerful probe for fundamental correlation physics in perovskites, especially with implications that polaron correlation may be critical for perovskites.

As illustrated in Fig. 1a, the ground state of MAPbI$_3$ at room temperature exhibit the bandgap E$_g$ $\sim$1.51 eV that is strongly influenced by the orientations and couplings of the MA$^{+}$ cations \cite{Jie, Artem} and octahedral PbI$_6$ halide cages \cite{Omer}. 
The pump-induced band edge coherent oscillations are probed by a delayed, white light continuum pulse ranging from 735 nm to 785 nm (inset, a 2D false-colour plot) after THz pump.
Here the wavelength dependent, transient differential reflectivity $\Delta$R/R in a MAPbI$_3$ single crystal is recorded as a function of pump-probe delays $\Delta t_{pp}$ after THz pumping, E$_{THz}$ =938 kV/cm at room temperature.
The observed $\Delta$R/R signals exhibit a sign reversal above and below 755 nm that is consistent with the known band gap shift and Pauli blocking nonlinearities from the presence of transient carrier population. The $\Delta$R/R dynamics measures free charge carrier decay. 
Most intriguingly, a pronounced oscillatory behavior is clearly visible and superimposes on the amplitude decay profile, as shown in Fig. 2b for three representative traces at wavelengths 755 nm (blue), 745nm (green) and 770 nm (black) selected from the 2D plot (dashed lines). The coherent oscillations persist at least the first $\sim$2 ps and appear to dephase together with the amplitude decay, e.g., as seen in the exponential decay fitting for the 770 nm trace (red solid line). 

Fig. 1c shows the origin of the periodic oscillations as the LO phonon coherence at $\omega_{\mathrm{LO}}$ induced by THz pump.
As an example, we isolate an oscillatory component $\Delta$R$_{osc}$/R at 770 nm (blue, inset of Fig. 1c) by removing the best fitted exponential amplitude decay ($\tau_{\mathrm{exp}}$, red solid line, Fig. 1b) from the $\Delta$R/R signal (black, Fig. 1b). 
The $\Delta$R$_{osc}$/R dynamics clearly show quantum beatings of multiple modes. 
The Fourier transformation (FT) spectra of $\Delta$R$_{osc}$/R dynamics display these two modes, one pronounced peak centered at $\sim$4 THz ($\#$2), which matches the LO phonon $\omega_{\mathrm{LO}}$ frequency \cite{Nagai} that is discussed in the context of for the polaron formation \cite{Miyata}, and the other secondary peak at $\sim$2 THz ($\#$1), which is consistent to the TO phonon mode $\omega_{\mathrm{TO}}$ seen in ultrafast THz and infrared absorption of MAPbI$_3$\cite{Luo,Leguy}. 
The $\omega_{\mathrm{LO}}$ mode from coherent dynamics shows a full width at half-maximum (FWHM) $\sim$2 THz (blue line, Fig. 1c) which is consistent to the lifetime-limited dephasing $\sim$1 ps measured in time domain (black, Fig. 1b).  
Note that both the $\omega_{\mathrm{LO}}$ and $\omega_{\mathrm{TO}}$ modes (marked by arrows) are also clearly visible in the FT spectra of the $\Delta$R/R raw data (black line) without removing the amplitude decay component, as shown in Fig. 1c. The exponential subtraction helps suppress the low frequency ($<$1THz) background in $\Delta$R$_{osc}$/R spectra. Moreover, much stronger nonlinearity is seen using 10 times higher THz pumping field in comparison to the second order nonlinearity observed in previous THz pump study\cite{Kim}. We emphasize two points below. First, the second order, electro-absorption mechanism dictates $\Delta$R/R signals to be scaled with the square of the incident THz E-field during the pulse. In contrast, the THz pump-induced $\Delta$R/R signal here cannot be scaled quadratically as discussed in supplementary (Fig. S3). Second, the 4 THz peaks observed don’t show any oscillating behavior associated with the electro-absorption mechanism \cite{Kim}, e.g., 770 nm (Fig. 1c) and 745 nm probe (Fig S4, supplementary) show the same behavior. 
Third, the pump-induced $\Delta$R/R after the pump pulse as a function of THz E-field which show a threshold behavior (Fig. S5, supplementary). 

\begin{figure*}[!tbp]
	\includegraphics[scale=0.6]{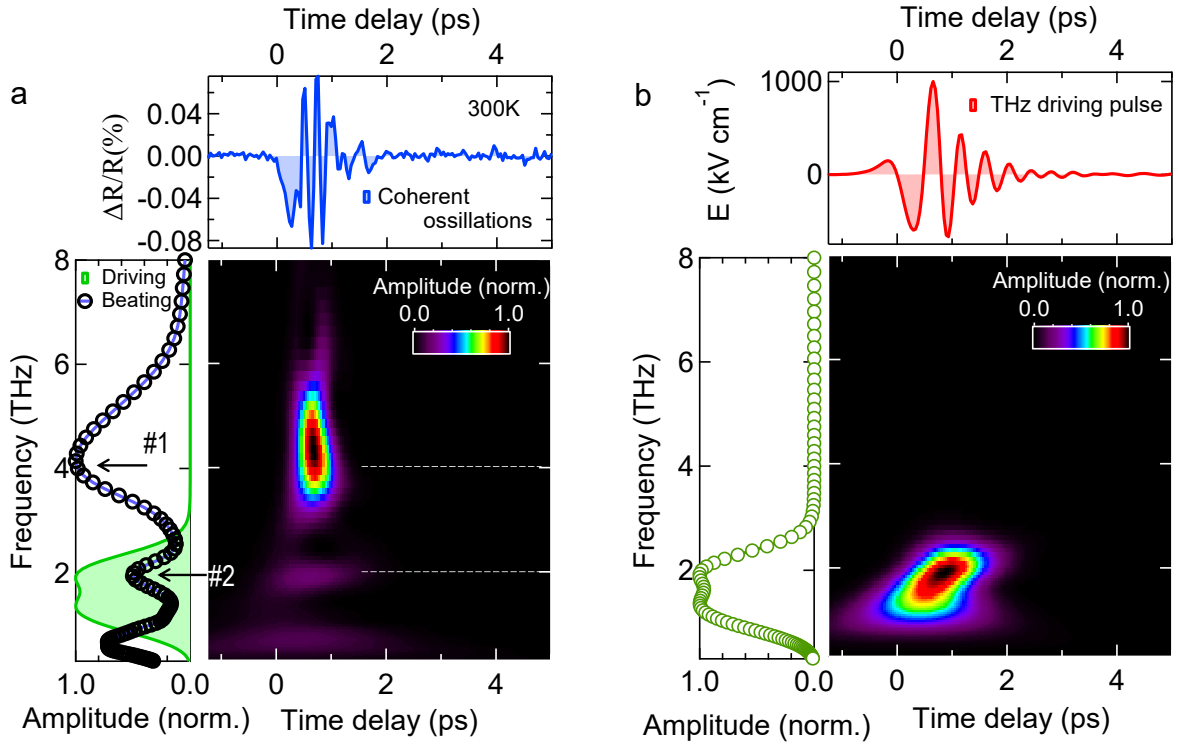}
	\caption{\textbf{THz {\em$\omega_{\mathrm{LO}}$ mode-specific} electronic band coupling revealed by intense THz pump-induced quantum beat dynamics and spectra at room temperature.} Continuous wavelet transform of the oscillatory component $\Delta$R$_{osc}$/R at probe wavelength 770 nm (a) and THz driving field (b). The top panels in (a) and (b) show the $\Delta$R$_{osc}$/R (blue line) and THz driving electric field (red line) in time domain. The time-integrated power spectra from time-frequency analysis of the 770 nm (blue circles) and THz field (green circles) are shown on the left in (a) and (b).}
	\label{Fig1}
\end{figure*}   

Note that the $\omega_{\mathrm{LO}}$ coherence and dephasing times, revealed in the time domain here by single-cycle intense THz pump, is close to the previous optical pump results that reveal the polaron mode $\sim$3.7 THz temporal oscillation \cite{McGill} by detecting the differential frequency between weak THz probe and polaron collective emission. 
Note that all these results are consistent with the formation time $\sim$1ps of large polarons \cite{Simon}. Polarization dephasing can lead to conversion of LO phonon coherence into population of polarons 
associated with the establishment of dynamic screening of Coulomb interactions. This indicates that the correlated electron-lattice interaction may contribute the observed signals.   

We transform the oscillatory component $\Delta$R$_{osc}$/R to time-frequency domain (Fig. 2a) and compare it with the THz driving pulse (Fig. 2b), in order to further substantiate the observation of the room temperature $\omega_{\mathrm{LO}}$ polaronic coupling,. Fig. 2a plots the Continuous Wavelet Transforms (CWT chronograms) of $\Delta$R$_{osc}$/R at probe wavelength of 770 nm for $\Delta$t$_{pp}$ up to 5 ps (see methods).
The distinct $\omega_{LO}$ mode is clearly visible in the time-integrated power spectrum (blue, side panel). In stark contrast to the $\omega_{\mathrm{LO}}$ quantum oscillations, THz driving field in Fig. 2b exhibits much slower oscillations (red line, top panel) with spectrum centered at $\sim$6~\,meV or 1.5 THz (green, side panel), far below the $\omega_{LO}$ frequency. This excludes the pump electric field oscillation as the origin of $\Delta$ R$_{osc}$/R coherent dynamics that exhibit the faster $\omega_{\mathrm{LO}}$ dynamics from the polaronic coherent coupling. Instead, the direct coupling between the THz pump (green shade) and optically-active, $\omega_{\mathrm{TO}}$ phonon (black circles), both located $\sim$2 THz, gives rise to the infrared active mode seen in Fig. 2a.  
In addition, the CWT plot also allows the determination of subcycle build-up of the $\omega_{\mathrm{LO}}$ mode during the THz pumping with dephasing time $\sim$1 ps. 
Although the dephasing time measured here is in the limit of THz pump pulse envelope, it should be noted THz pump effects are mostly on the first cycle of THz pump field instead of pulse envelope due in part to the strong threshold behavior associated with the tunneling ionization process. In addition, the dephasing time measured by single-cycle THz-driven quantum beats is consistent with the reported sub-ps polaron formation\cite{McGill,Simon} and the rotation correlation time of organic cations\cite{Guy,Nathaniel}. 

\begin{figure*}[!tbp]
	\includegraphics[scale=0.6]{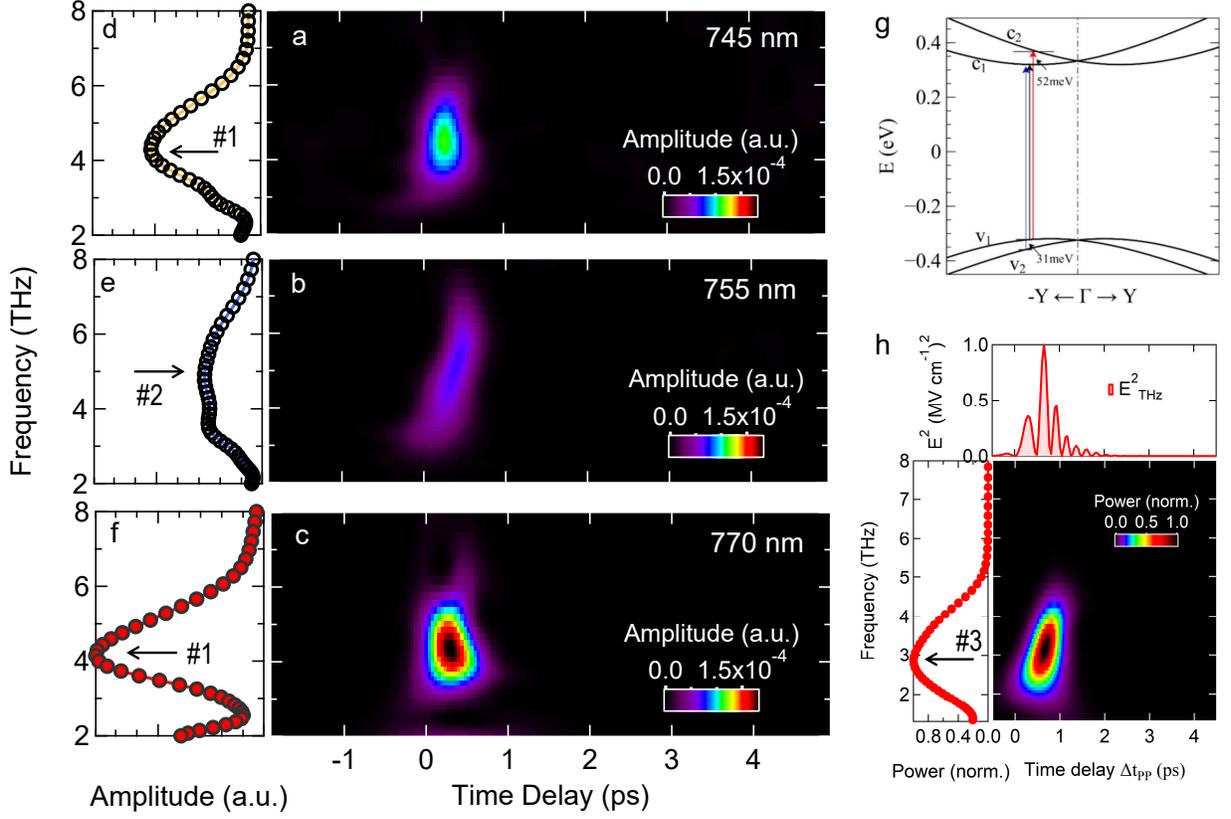}
	\caption{\textbf{Selective THz periodic modulation of the band edge electronic states from their coupling to the {\em$\omega_{\mathrm{LO}}$ mode-specific} coherence and dynamics at room temperature.} Continuous wavelet transform of oscillatory components $\Delta$R$_{osc}$/R at probe wavelength 745 nm (a), 755 nm (b), and 770 nm (c) at room temperature. They correspond to the three electric dipole-allowed inter-band transitions involving band extrema as discussed in supplementary (Fig. S1). (d)-(f) The time-integrated power spectra from time frequency-analysis of the 745 nm (orange circles), 755 nm (blue circles), and 770 nm (red circles) traces. The raw $\Delta$R$_{osc}$/R data used in time domain are shown in supplementary (Fig. S2). (g)  Band structure of MAPbI$_3$ calculated at DFT-PBE level near zero Fermi energy. Y point is defined as (0,\ 0.5,\ 0) in the conventional basis. (h) The time-integrated power spectra from time frequency analysis of the squared THz field E$^2_{THz}$ (red circles).}
	\label{Fig1}
\end{figure*}    

Fig. 3 demonstrates {\em mode-selective coupling} of the $\omega_{\mathrm{LO}}$ quantum coherence to different band edge states at room temperature. 	
We examine closely three CWT time-frequency spectral for the band edge states in Figs. 3a-3c at probe wavelengths 745 nm (yellow), 755 nm (blue) and 770 nm (red), respectively, up to $\Delta$t$_{pp}=$4.9 ps. These positions are determined by the locations of the positive and negative extrema as well as the zero point of the $\Delta$R/R transient spectra and don't change with time delays (Inset, Fig. 1b). 	
All plots show the notable, 4 THz peak yet with distinctly different spectral power intensity indicative of a selective coupling to the $\omega_{\mathrm{LO}}$ coherent lattice oscillation for the different band edge states.  
First, stronger 4 THz peaks appear on two sides at 745 nm and 770 nm probes, respectively, above (Fig. 3a) and below (Fig. 3c) the middle 755 nm one. For example, coupling strength of the 755 nm (Fig. 3b) probe at $\sim$ 4THz is suppressed, $\sim$3 times weaker than 770 nm (Fig. 3f). 
Second, in comparison with the time integrated power spectra for electronic states on two sides (Fig. 3d and Fig. 3f), the 755 nm one (Fig. 3e) appears broadened, with the FWHM almost two times larger, asymmetric towards the high frequency side due to the appearance of extra spectral weight $\sim$5 THz. This mode is consistent with the assignment of a MA rotational mode at 165 cm$^{-1}$ for MAPbI$_3$ \cite{Nagai}. Here the 755 nm transition appears to connect more to the organic cation mode which suppresses the $\omega_{\mathrm{LO}}$ 4 THz mode, consistent with our conclusion.    
Third, the probe wavelengths match very well with three main inter-band transitions as show in Fig. 3g, i.e., $v_1\leftrightarrow c_1$, $v_2\leftrightarrow c_1$ and $v_1\leftrightarrow c_2$ where $c$ and $v$ are the valence and conduction bands of MAPbI$_3$. 
Fig. 3g plots our density functional theory calculations (see supplemental) that show the valence and conduction bands of MAPbI$_3$ near Fermi level along the -Y$\rightarrow\Gamma\rightarrow$ Y path, with Y$\equiv$(0,\ 0.5,\ 0) in the conventional basis. 
The band gap from DFT-GGA band structure calculation is known to be underestimated, with a theoretical value ($E_{c_1}-E_{v_1}$) of 0.64 eV compared with $\sim$1.6 eV experimentally. However, the relative positions of valence bands or conduction bands should be quite accurate for MAPbI$_3$, as it is a typical wide-band weakly correlated material. The vertical line indicates three electric dipole-allowed inter-band transitions involving band extrema. Remarkably, the differences of inter-band transition frequencies, $f_{v_1\leftrightarrow c_2}-f_{v_2\leftrightarrow c_1}$=21 meV and $f_{v_2\leftrightarrow c_1}-f_{v_1\leftrightarrow c_1}$=31 meV, match very well with the three probe frequencies of 745 nm (yellow), 755 nm (blue) and 770 nm (red), with spacings of 22 meV and 32 meV, respectively. This shows that the three wavelengths are probing three band edge states that can be selectively and periodically modulated assisted by polaronic coupling with the $\omega_{\mathrm{LO}}$ mode and coherence. Such LO phonon mode-selective coupling of the inter-band transition through polaronic coupling, as demonstrated experimentally, provides an extra knob to engineer electronic fine structure for applications at room temperature. 

\begin{figure*}[!tbp]
	\includegraphics[scale=0.55]{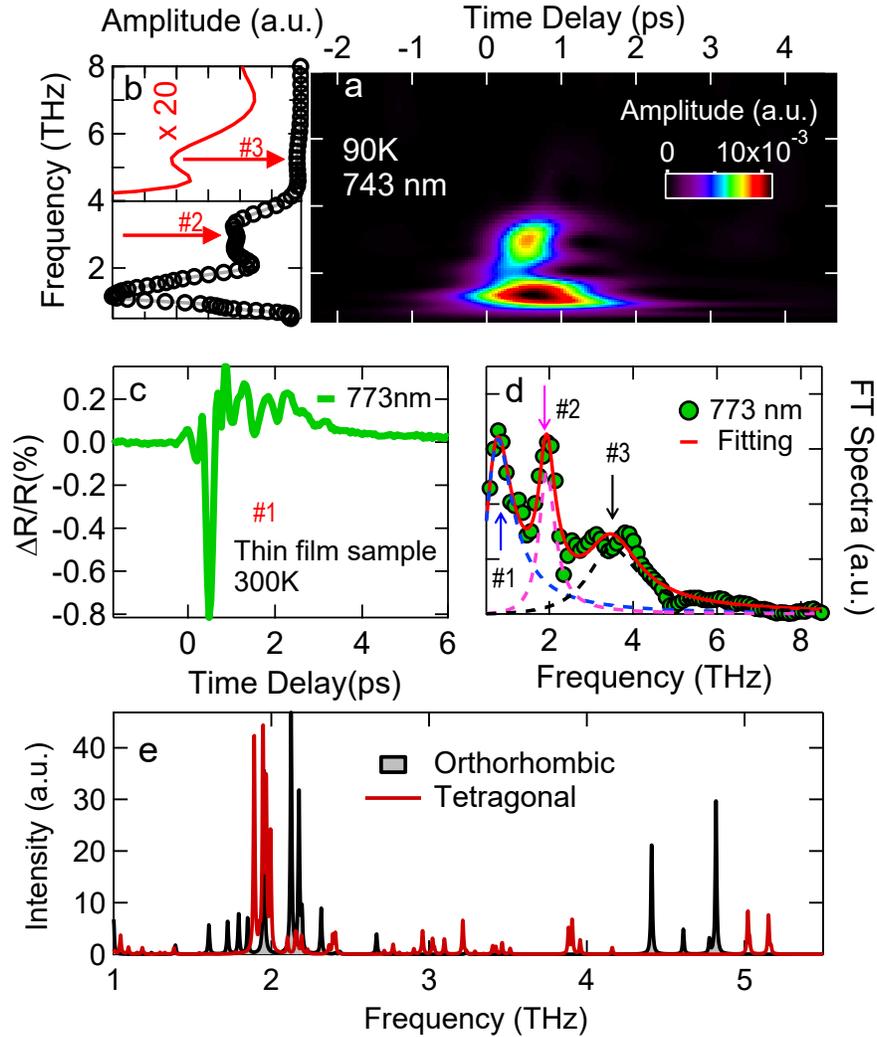}
	\caption{\textbf{Dynamic LO mode splitting and dynamics.} (a) Continuous wavelet transform of oscillatory component $\Delta$R$_{osc}$/R at probe wavelength 743 nm at E$_{THz}$=938 kV/cm and 90 K. (b) Time-integrated power spectrum from the time-frequency analysis of oscillatory component $\Delta$R$_{osc}$/R at probe wavelength 743 nm, with three peaks marked centered at $\sim$1 THz ($\#$1), 3.3 THz ($\#$2) and 4.9 THz ($\#$3, red solid line, $
		\times$20). (c) A representative $\Delta$R/R time trace of a MAPbI$_3$ thin film at room temperature induced by E$_{THz}$= 938 kV/cm at probe wavelength 773 nm. (d) The Fourier spectrum of the $\Delta$R/R time trace at probe wavelength 773 nm (green). Shown together is the line shape fitting (red line) obtained from 3 Lorentz oscillators (dashed colored lines) with center frequencies at 0.8 THz (blue), 1.9 THz (magenta) and 3.9 THz (purple). (e) Simulated THz phonon spectra convoluted with a 0.01 THz Lorentzian in tetragonal phase (red line) and orthorhombic phases (gray line). }
	\label{Fig1}
\end{figure*}

Comparing the spectra in Figs. 3d-3f and spectral-temporal analysis of squared THz pump pulse E$^2_{THz}$ in Fig. 3h, we can exclude the origin of the 4 THz mode from any second order electro-optical contribution from THz pumping. First, Fig. 3h, using the real pump pulse (Fig. 2b), shows a spectrum centered at $\sim$3 THz instead of the 4 THz as observed. This is consistent with the fact that the THz pump mostly centers at $\sim$1.5 THz not at 2 THz. Second, different center frequency and spectral broadening observed in the integrated power spectra between the side 770 nm (Fig. 3d) and the center 755 nm (Fig. 3e) cannot be explained by the electro-optical absorption mechanism. Third, we can also exclude the origin of the 4 THz mode from the electro-optical type effect from low temperature measurement discussed below.  

Distinctly different LO phonon modes emerge at low temperature signifies the salient roles of entropy and anharmonicity of MA$^{+}$ cation rotations. We analyze the influence of temperature on the polaron formation in MAPbI$_3$ single crystal, as shown in Fig. 4a, by plotting the CWT spectra of the oscillatory component $\Delta$R$_{osc}$/R at 90 K (the negative peak position at 743 nm as probe). This band has been shown to mostly couple to optically dark $\omega_{\mathrm{LO}}$ instead of IR-active MA$^{+}$ mode $\sim$5 THz shown in Fig. 3e. In contrast to the dominant $\omega_{\mathrm{LO}}$ coherent dynamics at room temperature in Fig. 3d, the quantum beat spectra largely lie below 4 THz centered at $\omega_\mathrm{1}$=1 THz (arrow $\#$1) and $\omega_\mathrm{2}$=3.3 THz (arrow $\#$2). These modes match well with the known TO phonons of PbI$_6$ perovskite cage\cite{Leguy,Brivio,Oscorio}. It is well established that MAPbI$_3$ perovskites go through a first-order phase transition across 160 K, i.e., transforming from room temperature tetragonal to low temperature orthorhombic phase. The deformation of PbI$_6$ cage below the phase transition temperature strongly restricts the rotation of organic cation.
As a consequence, we can attribute the absence of 4 THz mode at low temperature to the ``frozen” of organic molecular cation which transfer the spectral weight to low frequency modes. Most intriguingly, through careful examination of the power spectra in Fig. 4b, we observed a small peak at higher frequency, lying above 4 THz, centered at $\omega_\mathrm{3}\sim$5 THz (arrow $\#$3) in both original (black empty circles) and zoomed (red solid line, $\times$20) traces. This observation can be explained by the phonon split of the $\omega_{\mathrm{LO}}$ at 4 THz into two, i.e., the low frequency $\omega_\mathrm{2}$ and high frequency $\omega_\mathrm{3}$ modes due to coupling between MA$^{+}$ cation and the LO phonon. This is in excellent agreement with the recent linear THz absorption spectroscopy study that shows the 4 THz LO mode at room temperature splits into two modes, lying above the below at $\sim $3.3 THz and 4.9 THz respectively, in the low temperature orthorhombic phase\cite{Nagai}. Our results here provide the evidence of coherent excitation and detection of these modes. Moreover, comparing the quantum beat spectra at room (Fig.3) and low temperatures (Fig. 4a), we can exclude the origin of the 4 THz mode from any second harmonic generation contribution from THz pumping which is expected to be enhanced by decreasing temperature and without mode splitting. 

We further examine single-cycle THz-driven dynamics in device-relevant MAPbI$_3$ thin film samples at room temperature. We establish below the fundamental discovery of LO phonon quantum coherence, revealed in single crystals so far, also appears in perovskite films.   
Fig. 4c presents a representative $\Delta$R/R time trace at probe wavelength of 773 nm after THz excitation E$_{THz}$=938 kV/cm at room temperature. 
Similar to single crystal samples, we observe pronounced quantum oscillations in time domain lasting for few ps.  The FT spectrum of the raw data (green solid circles, Fig. 4d) reveals clearly three peaks, marked by arrows, including the high frequency one $\sim$4 THz indicative of the THz-driven $\omega_{\mathrm{LO}}$ quantum coherence. The polycrystalline grains in thin film samples is likely the main reason for slightly different mode positions and spectral weight distribution\cite{Reid,Shao}.   
The thin film spectrum can be well fitted by three Lorentz oscillators with center frequencies at 0.8 THz (blue), 1.9 THz (magenta) and 3.9 THz (purple). 
The first two modes are consistent with the infrared-active TO phonon modes commonly seen in thin film samples\cite{Luo} while the last one represents the observation of the previously-inaccessible $\omega_{\mathrm{LO}}$ mode that has large polaronic coupling and exists in both the MAPbI$_3$ single crystals and thin films. 

Finally, we discuss the temperature-dependent coupling between electrons and the LO phonon/organic MA cations in different structural phases observed in comparison with the First-principles modeling of IR phonon spectra at different phases controlled by temperatures.  
Tetragonal phase of MAPbI3 allows the fast rotation of the MA cations in the perovskite cage which causes the dynamic disorder and damped coherent oscillations at high temperature. These are among the most important factors responsible for reduction of the carrier mobility and diffusion length. At the low temperature orthorhombic phase the cations are restricted which leads to the characteristic mode splitting of the LO phonon observed in Fig. 4a. 
To put this on a strong footing, Fig. 4 provides the First-principles modelling of IR phonon spectra at different temperatures. 
We compare the theoretical zone-center IR phonon spectrum of the sample in the room temperature tetragonal phase with that in the low temperature orthorhombic phase. In consistent with experiment, high density of phonon modes close to 4 THz are clearly present at tetragonal phase. As the system transforms from tetragonal phase to orthorhombic phase with lowing temperature, the spectral weight near 4THz splits to upper and lower frequency modes, driven by the anharmonic phonon-phonon coupling effect through phase transition. This is consistent with our results from the observed coherent dynamics. At the low temperature phase, exciton plays a significant role as the binding energy becomes bigger than thermal energy As temperature rises, exciton binding energy decreases. At room temperature, the exciton becomes irrelevant in this bulk material. The free carrier (electron/hole)-phonon coupling can form polaron. The specific phonon modes, especially the 4 THz LO mode can strongly contribute to the large polaron formation. The simulation reveals the temperature-dependent phonon mode locking and phonon weight transfer in perovskites, which is critically depend on phonon-phonon coupling between the perovskite cage and organic cations. 

In summary, subcycle THz-driven electronic band edge modulations reveal the coherent LO dynamics. 
Our results demonstrate this technique as a powerful probe of collective modes in perovskites and provide compelling implications towards distinguishing polaronic correlation. We also reveal mode-elective modulation of electronic states and LO mode splitting that may have far reaching consequence of coherently enhanced photovoltaics, e.g., in analogy to the biological light-harvesting seemingly assisted by sub-ps electronic coherence \cite{Jean}. 
\\  

This work was supported by the U.S. Department of
Energy, Office of Basic Energy Science, Division of
Materials Sciences and Engineering (Ames Laboratory is
operated for the U.S. Department of Energy by Iowa State
University under Contract No. DE-AC02-07CH11358).
Sample development was supported by National Science Foundation DMR 1400432.
Terahertz spectroscopy instrument was supported 
by National Science Foundation EECS
1611454.

\noindent$^{\ast}$ jgwang@iastate.edu; jwang@ameslab.gov

\end{document}